\documentclass[12pt]{article}

\usepackage{amsmath,amssymb}
\usepackage{graphicx}

\makeatletter

\renewcommand\section{\@startsection{section}{1}{\z@}
                                   {-3.5ex \@plus -1ex \@minus -.2ex}
                                   {2.3ex \@plus .2ex}
                                   {\normalfont\large\bfseries}}
\renewcommand\subsection{\@startsection{subsection}{2}{\z@}
                                   {-3.25ex\@plus -1ex \@minus -.2ex}
                                   {1.5ex \@plus .2ex}
                                   {\normalfont\normalsize\bfseries}}
\renewcommand\subsubsection{\@startsection{subsubsection}{3}{\z@}
                                   {-3.25ex\@plus -1ex \@minus -.2ex}
                                   {1.5ex \@plus .2ex}
                                   {\normalfont\normalsize\bfseries}}
\renewcommand\paragraph{\@startsection{paragraph}{4}{\z@}
                                   {3.25ex \@plus1ex \@minus.2ex}
                                   {-1em}
                                   {\normalfont\normalsize\bfseries}}

\makeatother

\newcommand{\slambda}{S}

\newcommand{\be}{\begin{equation}}
\newcommand{\ee}{\end{equation}}
\newcommand{\bea}{\begin{eqnarray}}
\newcommand{\eea}{\end{eqnarray}}
\newcommand{\ba}{\begin{array}}
\newcommand{\ea}{\end{array}}

\newcommand{\id}{\hbox{1\kern-.27em l}}
\newcommand{\lb}{\langle}
\newcommand{\rb}{\rangle}
\newcommand{\half}{ {\textstyle \frac{1}{2}  } }

\newcommand{\al}{\alpha}
\newcommand{\ga}{\gamma}
\newcommand{\Ga}{\Gamma}
\newcommand{\bet}{\beta}

\newcommand{\de}{\delta}

\newcommand{\ep}{\epsilon}

\newcommand{\la}{\lambda}

\newcommand{\Om}{\Omega}

\newcommand{\tha}{\theta}

\newcommand{\bA}{\bar{A}}
\newcommand{\bD}{\bar{D}}

\newcommand{\bW}{\bar{W}}
\newcommand{\bd}{\bar{d}}
\newcommand{\bp}{\bar{p}}

\newcommand{\btha}{\bar{\theta}}
\newcommand{\bla}{\bar{\lambda}}

\newcommand{\cN}{\mathcal{N}}

\newcommand{\pa}{\partial}

\newcommand{\non}{\nonumber}

\newcommand{\SU}{\mathrm{SU}}
\newcommand{\SO}{\mathrm{SO}}

\newcommand{\U}{\mathrm{U}}

\newcommand{\ts}{\textstyle}

\begin{document}

\setcounter{footnote}{0}
\stepcounter{table}

{\small
\begin{flushright}
CERN-PH-TH/2005-169\\
\phantom{hep-th}
\end{flushright}
}

\vspace{-2mm}

\begin{center}
{\Large\sf Lower-dimensional pure-spinor superstrings}

\vskip 5mm

P. A. Grassi$^{\dag,\ddag,\S}$ and Niclas Wyllard$^{\S}$ 

\vskip 6mm
$^\dag$DISTA, Universit\`a del Piemonte Orientale, \\
Via Bellini 25/g 15100 Alessandria, Italy\\[1mm]

\vskip 3mm
$\ddag$ Centro Studi e Ricerche E. Fermi, \\
Compendio Viminale, I-00184, Roma, Italy\\[1mm] 

\vskip 3mm
$^\S$Department of Physics, Theory Division, CERN, \\
1211 Geneva 23, Switzerland

\vskip 3mm 

{\tt pgrassi,wyllard@cern.ch}
\end{center}

\begin{abstract}
\noindent We study to what extent it is possible to generalise Berkovits' 
pure-spinor construction in $d{=}10$ to lower dimensions. 
Using a suitable definition of a ``pure'' spinor in $d=4,6$, we propose  
models analogous to the $d=10$ pure-spinor superstring in these dimensions. 
Similar models in $d=2,3$ are also briefly discussed.
\end{abstract}

\setcounter{equation}{0}
\section{Introduction and summary}

One of the most important developments in 
superstring theory in recent years is the discovery of the pure-spinor 
superstring~\cite{Berkovits:2000a}. 
In this new formulation, Lorentz covariance and supersymmetry are 
manifest and quantisation can be achieved.   
Among the virtues of the pure-spinor formalism are that 
it is possible to calculate scattering amplitudes, 
derive effective field theories and to compute higher-derivative 
corrections  in a manifestly supersymmetric way, 
see e.g.~\cite{Berkovits:2004a,Anguelova:2004}. 
During the last few years the pure-spinor superstring  
has been studied and developed from several points of view, but 
further work is required in order to fully understand the basis of the model  
and its implications. 
For that reason, the study of similar but simpler models, based on the 
same structure, might shed some new light on the $d=10$ formulation. 
In this note we construct and study some such models. 

As is well known, at the classical level the superstring can be formulated 
in terms of different sigma models. One is the well-known RNS 
superstring. Another model, known in the literature as the Green-Schwarz (GS)
superstring~\cite{Green:1983}, 
has local worldsheet reparametrisation invariance   
and also a further symmetry, the so called $\kappa$-symmetry, 
which guarantees that 
target-space supersymmetry is implemented. 
However, no completely satisfactory way to quantise the GS model 
has been discovered to date. The only known way to quantise the GS superstring 
is in the light-cone gauge, but this approach has various drawbacks, 
e.g.~Lorentz covariance is not manifest and the calculation of 
general scattering amplitudes is problematic.

A way around the quantisation problems of the GS formulation 
is to  add to the theory the variable \cite{Siegel:1985}
$p_{\al}$, conjugate to the Grassmann-odd 
GS variable $\theta^{\al}$ (here $\al=1,\dots,16$) 
and in addition also add a Grassmann-even spinor 
ghost field~\cite{Berkovits:2000a} 
$\lambda^{\al}$ (and its conjugate $w_{\al}$). In this way, the worldsheet 
action for the superstring in a flat ten-dimensional supergravity background 
becomes a quantisable free action with manifest spacetime supersymmetry. 
It turns 
out that only if one assumes that the ghost field $\la^{\al}$ 
is constrained can one make the total central charge vanish and 
make the Lorentz current algebra have the right properties. 
Berkovits discovered~\cite{Berkovits:2000a} 
that the appropriate constraint that needs to be imposed is 
$\la^{\al} \ga^{m}_{\al\bet}\la^{\bet} =0$. The spinors $\la^{\al}$ 
satisfying this equation are called pure spinors.

The goal of this note is to investigate if it is possible to generalise 
Berkovits' construction for quantising the superstring 
in $d=10$ to lower dimensions and, in particular, to 
those dimensions where (classically) a GS model exists, i.e.~$d=(2),3,4,6$. 
By analogy with the pure-spinor formalism in $d=10$ we take the  
worldsheet ghost fields to involve a Grassmann-even spinor, $\la$, 
(and its conjugate momentum $w$) and look for suitable constraints on 
$\la$. We refer to the constrained spinor $\la$ as a pure spinor in any 
dimension.

Even though a complete light-cone quantisation of the GS superstring is 
only possible in $d=10$, in the particle limit the GS superstring becomes a 
superparticle theory~\cite{Brink:1981} and 
one can quantise the theory in the light-cone gauge  
in all the dimensions listed above. 
In the particle limit, only the lowest state of the spectrum survives 
and one can check that it describes an on-shell massless gauge multiplet 
(for open superstrings) or an on-shell  massless supergravity multiplet 
(for closed superstrings). 
Therefore, the counting 
of the degrees of freedom for the lowest multiplet is known and it 
can be checked if this counting matches the counting from a pure-spinor 
construction.  
In the next section, it will be shown how one can use this argument 
in reverse to obtain information about the required number of 
independent components of the 
pure spinor for each of the dimensions listed above. This gives a necessary 
condition on the pure spinor but it does not uniquely determine 
which constraint the pure spinor should satisfy. 

Pure spinors (in the sense of Cartan) 
can be defined in any dimension, but that definition turns out to be 
not quite what we want. 
One clue about which constraint to use comes from the fact that in $d=10$ 
the super-covariant derivatives 
satisfy $\{D_\al,D_{\beta}\} = \ga_{\al\beta}^m \pa_m$. In order for 
the BRST operator $Q_0 = \la^\al D_\al$ to be nilpotent one therefore requires 
$\la^\al \ga^m_{\al\beta}\la^\beta = 0$, i.e.~the pure-spinor constraint. 
This construction is related to the general framework of 
spinorial cohomology proposed in 
\cite{Cederwall:2001a,Cederwall:2001b} and can be carried out 
in any dimension. 
 In later sections it will be checked that if one defines the pure-spinor 
constraints in this way by starting from the algebra of super-covariant 
derivatives then, in any of the dimensions listed above, 
one finds that the pure spinor has the right number of independent components 
to ensure that the central charge is $c=0$. 
Furthermore, one finds that the pure-spinor constraints obtained in this way 
are such that they imply $k=1$ (Lorentz current algebra has level one) 
in all the dimensions listed above.
 
We should point out that ideas related to this work have appeared 
in various places in the literature, e.g.~in section 2.6 
of~\cite{Berkovits:2002d} where pure-spinor superparticle models were 
suggested. These models correspond to the particle limit of our models. 
Another related paper is~\cite{Movshev:2005} where more mathematical  
aspects were studied.

This note is organised as follows. In the next section we first show how 
a heuristic counting of massless degrees of freedom gives 
the required number of independent components of the pure spinors 
in various dimensions. 
Then, using a decomposition of spinors similar to the one 
in~\cite{Berkovits:2000c},  
we show that the pure-spinor constraints obtained from the 
algebra of super-covariant derivatives lead to the same number 
of independent components as was obtained from the heuristic counting. 
In section~\ref{N1} we present the worldsheet theories of 
the new pure-spinor models in $d=3,4,6$. 
Then in section~\ref{N2} we analyse the worldsheet conformal field theory 
of the pure-spinor models in $d=4,6$. In particular, we determine 
the central charge and calculate the current algebra of the Lorentz 
currents (more details of these calculations will be given in a 
separate publication \cite{Wyllard:2005}). In section~\ref{N3}, 
using a BRST operator analogous to the one proposed by Berkovits 
in $d=10$, we analyse the cohomology at the massless level and show that 
the field content is exactly that of the super-Yang-Mills vector multiplet 
(further results will be presented in \cite{Grassi:2005b}).  
Finally, in section \ref{N4} we present our conclusions and discuss some 
open problems. 


\section{Pure-spinor constraints in various dimensions}
\label{N0}

In this section we present the pure-spinor constraints satisfied by the 
ghost field, $\la$, in our models. We first give a heuristic counting 
argument for how many independent components the pure spinor should possess. 
We then present the pure-spinor constraints and check that they are such that 
the number of independent components agree with the heuristic counting.

\subsection{Counting of degrees of freedom}
\label{prosciutto}

The number of independent degrees of freedom that the pure-spinor fields 
should possess can be investigated by a simple heuristic counting based 
on reparametrisation invariance and on the number of independent 
components of spinors in various dimensions. 

If we denote  by $d_{b}$ the number of bosonic coordinates, 
by $d_{f}$ the number of fermionic (Grassmann odd) coordinates, and 
by $d_{ps}$ the number of independent components of the pure spinor. 
In order to have a ``critical'' (i.e. $c=0$) model we need 
\be
d_{b} - 2 d_{f} + 2 d_{ps} = 0\,.
\ee
Using the known GS worldsheet structure (which gives $d_b$ and $d_f$), 
we find that $d_{ps}$ has to be 11 in $d=10$, 5 in $d=6$, and 2 in $d=4$. 
For the $d=3$ case, one finds the peculiar result $d_{ps} = \half$.

To implement the Virasoro constraints, we assume that  
two fermionic degrees of freedom have to be used 
to remove the lightcone coordinates $x^{\pm}$. We then find  
\be\label{countA}
(d_{b} -2 ) - 2 \, (d_{f} - d_{ps} - 1) =0 \,.
\ee
The same counting as in (\ref{countA}) also gives us the number of 
independent  degrees of freedom for the lowest state 
of the spectrum. For example, using the GS formalism 
the number of fermionic fields after the $\kappa$-symmetry 
is taken into account is equal to $2\,(d_{f} - d_{ps} -1)$. 
The above counting tells us that in $d=10$ there are $8$ bosonic massless 
degrees of freedom and $8$ fermionic ones, in agreement with the (on-shell) 
spectrum of $d=10$ super-Yang-Mills. 
In $d=6$, the above counting gives $4$ bosons plus $4$ fermions, and so on. 
However, in lower dimensions knowledge of the number of independent 
degrees of freedom alone is not sufficient to unambiguously infer 
to which supermultiplet they belong.


\subsection{The pure-spinor constraints}

As discussed in the introduction, in $d=10$ one can view the pure-spinor 
constraint as arising from the requirement that the BRST operator of 
the form $Q_0 = \la D$, where $D$ is 
the super-covariant derivative, is nilpotent. 
Guided by this fact and the structure of the algebra of super-covariant 
derivatives in lower dimensions  (see appendix \ref{appFlat}) we choose to 
impose the following ``pure-spinor'' constraints in $d=4,6$
\be \label{pure-spinor}
\ba{lcl}
\lambda \Gamma^{m} \lambda = 0 \,,  && \quad(d=4) \\
\ep^{IJ} \lambda_I \gamma^{m} \lambda_J = 0 \,. && \quad(d=6)    
\ea
\ee
In $d=4$ $\la^\al$ is a Dirac spinor 
and $\Gamma^m_{\al\beta}$ are the 
(symmetric) 
$4{\times}4$ gamma matrices, whereas in $d=6$ $\la^\al_I$ ($I=1,2$) 
is a doublet of Weyl spinors, and $\gamma^{m}_{\al\beta}$ are the 
(antisymmetric) $4 {\times} 4$ 
off--diagonal blocks (``Pauli matrices'') in the Weyl representation of 
the 
$8 {\times} 8$ six--dimensional gamma matrices $\Gamma^m$. 
Note that the above  $d=6$ condition is not the conventional 
pure-spinor condition (which is solved by a Weyl spinor). However, as 
confusion is unlikely to arise we refer to the conditions 
(\ref{pure-spinor}) 
as pure-spinor conditions throughout this note.
A convenient way to write the 
$d=4$ pure-spinor condition is to use 
two-component Weyl notation: $\la^a \bla^{\dot{a}}=0$ (here $a,\dot{a}=1,2$).
In $d=3$, a possible pure-spinor condition is given by 
the vector condition ($\al,\beta = 1,2$)
\be\label{tre}
\la \Ga^m \la = 0 \quad \Leftrightarrow \quad \la^{\al} \la^{\bet} = 0 \,. 
 \qquad   (d=3) 
\ee
This condition has to be treated with care. Since the constraint is 
reducible it imposes less than 3 constraints on the spinor. 


\subsection{Solving the pure-spinor constraints}

In $d=10$, one can decompose the pure spinor as~\cite{Berkovits:2000c}   
$\la^{\al} = (S^{a},S^{\dot a})$ where $a,\dot a =1, \dots, 8$.  
Using this decomposition, the pure-spinor constraint $\la \ga^{m} \la =0$ 
can be written 
\be
S^{a} S^{b} \delta_{ab} =0\,, \qquad
S^{a} S^{\dot b} \sigma^{I}_{a \dot b} =0\,, \qquad
S^{\dot a} S^{\dot b} \delta_{\dot a \dot b} =0\,. 
\ee
The first constraint reduces the number of independent 
components of $S^{a}$ to $7$. 
The other constraints are solved by an infinite  number 
of gauge symmetries which reduce the number of components of $S^{\dot a}$ from 
$8$ down to $4$.\footnote{Let us also mention that in order 
to substantiate the argument given in subsection \ref{prosciutto} on 
the use of reparametrisation invariance to count the degrees of freedom, 
one can relax the scalar constraint 
$S^{a} S^{b} \delta_{ab} =0$ by adding it to the BRST operator. 
For consistency, one is then forced to also add 
a further piece to the BRST operator of the form $c \Pi$, where $c$ 
is a Grassmann-odd ghost field and $\Pi$ is 
the momentum along a specific direction. This new piece imposes 
a light-cone constraint on the coordinates $x^{m}$ and plays the role of  
the Virasoro constraint.}

In the $d=6$ case, one can use the decomposition  
$\la^{\al}_{I} = (\slambda^{a}_{I}, \slambda^{\dot a}_{I})$ 
($a,\dot a =1, \dots, 2$). The above 
pure-spinor constraint then decomposes into   
\be\label{seiPS}
\slambda^a_I \slambda^b_J \ep_{ab} \epsilon^{IJ} =0\,, \qquad
\slambda^{\dot a} _I \slambda^{\dot b}_J 
\ep_{\dot a\dot b} \ep^{IJ} =0\,, \qquad
\slambda^a_I \slambda^{\dot b}_J \ep^{IJ} =0 \,.
\ee
The second and third conditions are solved by introducing 
a new set of fields $\rho^{\dot b}_{(0) \, a}$ and by setting 
\be
\slambda^{\dot b}_{I} = \rho^{\dot b}_{(0) \, a} \slambda^{a}_{I}\,.
\ee
It can easily be checked that this solves the constraints above if the 
first constraint is satisfied. The new field $\rho^{\dot b}_{(0)\, a}$ 
has 4 degrees of freedom, but there are gauge symmetries to take 
into account. Indeed, we have that 
\be
\delta \rho^{\dot b}_{(n) \, a} 
=  \ep^{IJ} \ep_{a b} \,  \eta^{\dot b}_{(n) \, I}  \slambda^{b}_{J}\,, ~~~~~
\delta \eta^{\dot b}_{(n)\, I} 
= \rho^{\dot b}_{(n+1)\, a} \slambda^{a}_{I}\,, 
~~~~~~ 
\dots 
\ee
which leads to an infinite number of fields $\rho^{\dot \bet}_{(n) \, a}$ and 
$\eta^{\dot b}_{(n)\, I}$. This reduces the number of components  
of $\slambda^{\dot a}_{I}$ from 4 to 2. 
The first constraint in  (\ref{seiPS}) reduces the number 
of independent components of $\slambda^{a}_{I}$ down to 3. Hence, the 
total number of independent degrees of freedom is $5$. 

In the $d=4$ case we use the decomposition  
$\la^{\al} = (\slambda^{\pm}, \bar{\slambda}^\pm)$.  
The pure-spinor constraint then decomposes into  
\be
\slambda^{+} \bar \slambda^{+}=0\,, ~~~~
\slambda^{+} \bar \slambda^{-}=0\,, ~~~~
\slambda^{-} \bar \slambda^{+}=0\,, ~~~~
\slambda^{-} \bar \slambda^{-}=0\,. ~~~~
\ee
In analogy with the $d=10$ and $d=6$ cases, we solve the 
last 3 constraints by setting 
\be
\slambda^{-} = \rho_{(0)} \slambda^{+}\,, ~~~~~
\bar\slambda^{-} = \bar\rho_{(0)} \bar\slambda^{+}\,,  
\ee
which are defined up to the (infinite) gauge symmetries 
\be
\delta  \rho_{(n)} = \eta_{(n)} \slambda^{+}\,, ~~~
\delta  \bar\rho_{(n)} = \bar\eta_{(n)} \bar\slambda^{+}\,, ~~~
\delta \eta_{(n)} = \rho_{(n+1)} \slambda^{+}\,, ~~~
\delta \bar\eta_{(n)} = \bar\rho_{(n+1)} \bar\slambda^{+}.
\ee
This reduces the pure spinors $\slambda^{-}, \bar\slambda^{-}$ to 
just one by means of an infinite number of fields. 
In addition, there is the first constraint 
$\slambda^{+} \bar \slambda^{+}=0$, which 
also reduces the number of components by one, so the final solution is 
just 2 degrees of freedom. 

We have shown that the number of independent components of the pure spinor 
is  two in $d=4$ and five in $d=6$ (and eleven in $d=10$). 
In other words, we see that in $d=2n$ the pure-spinor condition eliminates $n$ 
components from $\la$. 
An alternative way to obtain this result is to solve the pure-spinor 
constraints by temporarily breaking the $\SO(2n)$ (Wick rotated) 
Lorentz invariance to $\U(n) \simeq \SU(n) {\times} \U(1)$. 
This approach is discussed in more detail in~\cite{Wyllard:2005}.

We end this section by briefly discussing the $d=3$ case. 
If we use the decomposition $\la^{\al} = (\slambda^{\pm})$, 
then the pure-spinor constraint becomes 
\be
\slambda^{+} \slambda^{+} =0\,, ~~~~~ \slambda^{+}\slambda^{-} =0\,, ~~~~~
\slambda^{-} \slambda^{-}=0\,.
\ee
In analogy with the above situation, we solve the second and 
the last constraint by setting 
\be
\slambda^{-} = \rho_{(0)} \slambda^{+}\,.
\ee
This is defined up to gauge symmetries given by 
\be
\delta \rho_{(n)} = \eta_{(n)} \slambda^{+}\,, ~~~~
\delta \eta_{(n)} = \rho_{(n+1)} \slambda^{+} \,.
\ee
This infinite gauge symmetry reduces the number of components of the 
``pure spinor'' $\slambda^{-}$ to ``1/2 degrees of freedom''. 
It is not entirely clear how to interpret this result, but it 
will be shown in section \ref{N3} that the BRST cohomology makes sense 
and we explicitly compute the cohomology at the massless level. 
Further analysis and results will be presented elsewhere \cite{Grassi:2005b}. 


\setcounter{equation}{0}
\section{The $d=3,4,6$ pure-spinor models}
\label{N1}

In this section we describe the worldsheet theories 
of the $d=3,4,6$ pure-spinor models in a flat supergravity background. 
For simplicity we only write the left-moving sector explicitly.

For the case of $\cN=1$ supersymmetry in $d=4$ we choose 
(following Siegel \cite{Siegel:1985}) the 
left--moving (holomorphic) ``matter'' worldsheet fields 
to be  $(x^{m}, \theta^\al,p_\al)$, where $\theta^\al$ is a 
four--component 
Dirac spinor, and $p_\al$ is its conjugate momentum ($\al=1,\ldots4$). 
The Dirac spinor $\tha^\al$ can be decomposed 
into a Weyl spinor, $\tha^a$, and an anti-Weyl spinor, $\btha^{\dot{a}}$. 
Similarly, $p_\al$ can be decomposed into $p_a$, and $\bp_{\dot{a}}$.

For the case of $\cN=(1,0)$ supersymmetry in $d{=}6$ we choose 
the left--moving (holomorphic) matter  
worldsheet fields to be $(x^{m}, \theta^\al_I,p_\al^I)$, where $\theta^\al_I$ 
is a doublet ($I=1,2$) of four--component Weyl spinors, and $p_\al^I$ 
are their conjugate momenta. 

For the case of $\cN=1$ supersymmetry in $d=3$ 
we choose the left--moving matter worldsheet fields 
to be $(x^{m}, \theta^{\al},p_{\al})$ where $\theta^{\al}$ is a 
two-component Majorana spinor, and $p_{\al}$ is its conjugate momentum 
($\al=1,2$). The Dirac gamma matrices $\Ga^{m}_{\al\bet}$ are 
symmetric and any symmetric bispinor $A^{\al\bet}$ 
is proportional to the gamma matrix itself 
$A^{\al\bet} = \Ga^{\al\bet}_{m} A^{m}$.

The worldsheet actions for the left-moving modes (in the conformal gauge) 
have the following form  
\bea \label{actionA}
&& S^{d=3}_{\rm left-moving} = \int d^{2}z \left( \eta_{mn} 
\partial x^{m} \bar\partial x^{n} 
+ p_{\al} \bar\partial \theta^{\al} + w_{\al} \bar\partial \la^{\al}
\right), \nonumber \\
&& S^{d=4}_{\rm left-moving} = \int d^{2}z \left( \eta_{mn} 
\partial x^{m} \bar\partial x^{n} 
+ p_{\al} \bar\partial \theta^{\al} + w_{\al} \bar\partial \la^{\al}
\right), \non \\
&& S^{d=6}_{\rm left-moving} = \int d^{2}z \left( \eta_{mn} 
\partial x^{m} \bar\partial x^{n} 
+ p^{I}_{\al} \bar\partial \theta^{\al}_{I} 
+ w^{I}_{\al} \bar\partial \la^{\al}_{I}
\right). 
\eea
Note that the third term in each of these actions is written as a covariant 
expression, but one has to take into account that the pure spinors 
satisfy a constraint. The pure-spinor constraint implies that 
the action is invariant under the gauge symmetries
\be
\ba{ll}
\delta w_{\al} = \Lambda_{\al\bet} \la^{\bet}\,, & \qquad (d=3)  \\
\delta w_{\al} = \Lambda_{m} \Ga^{m}_{\al\bet} \la^{\bet}\,, & \qquad 
(d=4) \\
\delta w_{\al}^{I} = \Lambda_{m} \ep^{IJ} \gamma^{m} \la_{J}^{\bet}\,. &
 \qquad (d=6)
\ea \label{gaugeA}
\ee
Some progress toward understanding these gauge symmetries has been made 
in~\cite{Grassi:2001} 
by adding new ghost fields. 
However, for the rest of the paper, 
we do not introduce any further degrees of freedom. 

The only consistent way to study the BRST cohomology is 
to restrict the functional space consisting of 
pure spinors and their conjugates to the space consisting of 
only gauge invariant combinations. 
Using the gauge symmetries (\ref{gaugeA}) one can easily check that 
the number of independent $w_{\al}$ and $w_{\al}^{I}$'s is the 
same as the number independent components of the pure spinors. 
 
Notice that in order to be able to calculate correlation functions involving 
the pure-spinor fields using free field technology one has to fix the gauge 
to break the symmetry (\ref{gaugeA}). 
This can be done in several ways, for instance 
by selecting a suitable representation of the pure spinors and choosing the 
gauge where the only non-vanishing components of $w_{\al}$ and 
$w_{\al}^{I}$ are exactly the variables conjugate to the independent 
components  of the pure spinors. Computing correlation functions among 
gauge invariant operators, the details of the gauge fixings are irrelevant. 
An alternative method which does not rely on any particular gauge has been developed in \cite{Berkovits:2005}. 

It can easily be checked that the actions (\ref{actionA}) are 
Lorentz invariant and invariant under supersymmetry. Notice that 
one of the main difference with GS superstrings is 
the fact that the actions can be written in terms of supersymmetric 
variables without introducing a Wess-Zumino term. 
Up to terms that vanish by the equations of motion, 
the supersymmetric variables are given by 
$\pa \tha^\al$ ($d=4$), $\pa \tha_I^\al$ ($d=6$) as well as
\be
\ba{ll}
 d_\al = p_\al - \half (\Ga_m \tha)_\al \pa x^ m 
- \frac{1}{8} (\Ga_m \tha)_\al (\tha \Ga^m \pa \tha) \,,
& \quad (d=4)  \\[4pt] 
d^I_\al = p^I_\al - \half \ep^{IJ} (\ga_m \tha_J)_\al \pa x^ m 
- \frac{1}{8} \ep^{IJ} (\ga_m \tha_J)_\al 
\ep^{KL}(\tha_K \ga^m \pa \tha_L) \,,  & \quad (d=6)
\ea
\ee
and also
\be
\ba{ll}
\Pi^m = \pa x^m + \half \tha \Ga^m \pa \tha \,, &\qquad (d=4)  \\[3pt]
\Pi^m = \pa x^m + \half \ep^{IJ} (\tha_I \ga^m \pa \tha_J) \,. & \qquad (d=6)
\ea
\ee
The $d=3$ supersymmetric variables can be 
constructed in an analogous way. 
However, since they will not be used in this paper, we do not present 
the explicit formul\ae{} here. 


\setcounter{equation}{0}
\section{The pure-spinor conformal field theory in $d=4,6$}
\label{N2}

In this section we discuss the worldsheet conformal field theory for the 
$d=4,6$ pure-spinor models. Additional details will appear 
in \cite{Wyllard:2005}. 

The free fields $x^m$, $\tha$ and $p$ have the standard OPEs 
(in units where $\alpha'=2$),
\be
\ba{rcll}
 x^{m} (y, \bar{y}) x^{n} (z, \bar{z}) &\sim& -  \eta^{mn} \log
\left| y - z \right|^{2} \,, & \qquad(d=4,6)  \\
 p_{\alpha} (y) \theta^{\beta} (z) &\sim&
\frac{\delta_{\alpha}^{\beta}}{y-z}  \,, & \qquad(d=4)  \\
  p^I_{\alpha} (y) \theta_J^{\beta} (z) &\sim&
\frac{\delta^I_J\delta_{\alpha}^{\beta}}{y-z} \,. & \qquad (d=6) 
\ea
\ee
In terms of the supersymmetric variables given in the previous section 
we have the following OPEs for $d=4$ 
\bea
&& d_{\alpha} (y) d_{\beta} (z) \sim - \frac{1}{y - z}
\Ga_{\alpha\beta}^{m} \Pi_{m} (z)\,, \quad
d_{\alpha} (y) \Pi^{m} (z) \sim \frac{1}{y - z}
\left( \Ga^{m} \partial \tha \right)_{\alpha} (z)\,,\non \\
&& d_{\al}(y)\pa \theta^\beta(z) \sim \frac{1}{(y-z)^2}\de^\beta_\al \,,
\label{OPE-dd4}
\eea
and for $d=6$ 
\bea
&& d^I_{\alpha} (y) d^J_{\beta} (z) \sim - \frac{1}{y - z}
\ep^{IJ} \gamma_{\alpha\beta}^{m} \Pi_{m} (z)\,, \quad
d^I_{\alpha} (y) \Pi^{m} (z) \sim \frac{1}{y - z}
\ep^{IJ} \left( \gamma^{m} \partial \theta_I \right)_{\alpha} (z)\,,\non \\
&& d^I_{\al}(y)\pa \theta_J^\beta(z) \sim 
\frac{1}{(y-z)^2}\de^I_J\de^\beta_\al \,.
\label{OPE-dd6}
\eea
The ``matter'' part of the stress tensor is
\bea
T_{\rm mat} = -\half \pa x^m \pa x_m - p_\al \pa \tha^\al \,, 
&& \qquad (d=4) \non \\
T_{\rm mat} = -\half \pa x^m \pa x_m - p^I_\al \pa \tha_I^\al\,. 
&& \qquad (d=6)
\eea
From these expressions we see that the $x^m$ CFT has the standard
$c=d$ central charge, while the $(p,\theta)$ CFT has central charge 
$c=-8$ ($d=4$) and $c=-16$ ($d=6$). For the total central charge 
to vanish, the ghost CFT has to have $c=4$ ($d=4$) 
and $c=10$ ($d=6$). 

As in $d=10$, one can construct the manifestly 
$\SO(2n)$ Lorentz-covariant quantities 
\be
\ba{llll}
& N^{mn} = \half w \Gamma^{mn} \lambda \,, & \qquad 
\partial h = \half w \lambda \,, &  \qquad (d=4)  \\[2pt]
& N^{mn} = \half w^I \gamma^{mn} \lambda_I \,, &\qquad 
\partial h = \half w^I \lambda_I \,. & \qquad (d=6)
\ea
\ee
It can be shown \cite{Wyllard:2005} that, in $\SO(2n)$ covariant notation, 
the OPEs involving $N^{mn}$ and $\la$ take the form 
\bea \label{NNOPE}
N^{mn} (y) \lambda^\al (z) &\sim& \frac{1}{2}\ \frac{1}{y - z}
\left( \Gamma^{mn} \la \right)^\al(z) \,, \qquad \quad \; (d=4) \non \\[3pt]
N^{mn} (y) \lambda_I^{\alpha} (z) &\sim &\frac{1}{2}\ \frac{1}{y - z}
{\left( \gamma^{mn} \right)^{\alpha}}_{\beta} \, \lambda_I^{\beta} (z)
\,, \qquad \;\; (d=6) \\[3pt]
N^{p q } (y) N^{mn} (z) &\sim& \frac{ 
\eta^{p m} N^{q  n} (z) {-} \eta^{q m} N^{p n} (z) 
- (\mbox{\footnotesize $m$} \leftrightarrow \mbox{\footnotesize $n$} )}{y 
- z}  -
(n-2) \frac{\eta^{p n} \eta^{q m} {-} \eta^{p m}
\eta^{q n}}{(y - z)^{2}} \,. \non
\eea
where $n$ takes the values $2,3$ ($n=5$ corresponds to the $d=10$ case).
From the last equation in (\ref{NNOPE}) we see that the $N^{mn}$'s 
form an $\SO(2n)$ current algebra with level $k=2-n$.

In $d=4,6$ (as in $d=10$), $\pa h$ has no singular 
OPEs with $N^{mn}$ and satisfies 
\begin{equation}\label{extra1}
h(y) h(z) \sim - \log \left( y - z \right), \quad \partial h(y)
\lambda(z) \sim \frac{1}{2}\ \frac{1}{y-z} \lambda(z)\,.
\end{equation}

In comparison, the OPEs involving the $(p,\tha)$ Lorentz currents, 
$M^{mn} = -\frac{1}{2} p \Gamma^{mn} \theta$ ($d=4$) and  
$M^{mn} = -\frac{1}{2} p^I \gamma^{mn} \theta_I$ ($d=6$), take the form
\begin{eqnarray} \label{OPE-M}
M^{mn} (y) \theta^{\alpha} (z) &\sim&  \frac{1}{2}\ \frac{1}{y - z}
( \Gamma^{mn} \theta)^\al(z) \,,
\qquad \quad \;\;  (d=4)\\
M^{mn} (y) \theta_I^{\alpha} (z) &\sim&  \frac{1}{2}\ \frac{1}{y - z}
{\left( \gamma^{mn} \right)^{\alpha}}_{\beta} \, \theta_I^{\beta} (z) \,,
\qquad \;\; (d=6) \\
M^{p q} (y) M^{mn} (z) &\sim& 
\frac{ \eta^{p m } M^{q n} (z) {-} \eta^{ q m }
M^{p  n} (z) - 
( \mbox{\footnotesize $m$} \leftrightarrow  \mbox{\footnotesize $n$})
}{y - z} +
(n-1) \frac{\eta^{p n} \eta^{q m} {-} \eta^{p m}
\eta^{q n}}{(y - z)^{2}} \,. \non
\end{eqnarray}
Thus the $M^{mn}$'s form an $\SO(2n)$ current
algebra with level $k=n-1$. By combining the above results one finds that 
the total Lorentz current $L^{mn} = M^{mn} + N^{mn}$ satisfies the OPE
\begin{eqnarray}
\!\!\!\!\!\!\!\!\!\! 
L^{p q} (y) L^{mn} (z) \!\!\! &\sim& \!\!\! \frac{ \eta^{p m}
L^{q n}(z)  {-} \eta^{q m}
L^{p  n}(z) 
- (\mbox{\footnotesize $m$}\leftrightarrow \mbox{\footnotesize $n$}) }{y - z} 
+ \frac{\eta^{p n} 
\eta^{q m} {-} \eta^{p m}
\eta^{q n}}{(y - z)^{2}} \,,
\label{OPE-LL}
\end{eqnarray}
and thus forms an $\SO(2n)$ current algebra with level $k=1$ for all $n$, 
i.e.~independently of the dimension.

Using the covariant fields $N^{mn}$ and $\pa h$, {\it part of} 
the ghost stress 
tensor in $d=4,6,10$ can be written as \cite{Wyllard:2005}
\be\label{covst}
T_{N,\pa h} 
= - {\ts \frac{1}{4n}} N_{mn} N^{mn} - \half \left
( \partial h \right)^{2} + \frac{(n-1)}{2} \partial^{2} h \,. 
\ee
Above we said `part of' since in $d=6$ the stress tensor 
also contains the additional decoupled piece \cite{Wyllard:2005}
\be \label{Tuv}
T_{u,v} 
=  \partial u \pa v - \pa v^2 \,,
\ee
where $\pa u$ and $\pa v$ satisfy the OPE $\pa u(y)\,\pa v(z) \sim (y-z)^{-2}$.

Let us analyse the ghost stress tensor (\ref{covst})  
in more detail. The first piece
involves the ghost Lorentz currents, $N^{mn}$, and is a Sugawara
construction for an $\SO(2n)$ WZNW  model with level $k=2-n$. 
Indeed, recalling that the dual Coxeter number of $\SO(2n)$ is 
$g^\vee=2n-2$, we find\footnote{Due to our 
normalisation of the $N N$ OPE  the prefactor in front of 
$N_{mn}N^{mn}$  in (\ref{covst}) is $-\frac{1}{4(k+g^\vee)}$.  
To obtain the usual 
$+\frac{1}{2(k+g^\vee)}$ one would have to rescale the currents $N^{mn}$.} 
$2(g^\vee + k) = 2n$. Using standard formul\ae{} one finds that 
the central charge for the piece involving the ghost Lorentz currents is
\begin{equation}
c = \frac{k \dim \SO(2n)}{ k+g^\vee} = (2-n)(2n-1) \,.
\end{equation}
In (\ref{covst}) the second piece refers to a Coulomb gas, with a 
background
charge of $Q=(n-1)$, and consequently central charge 
$c = 1 + 3 Q^{2} = 1 + 3(n-1)^2$. 
Adding these two contributions gives $n^2-n + 2$. 
Taking into account that in $d=6$ the additional piece 
(\ref{Tuv}) has $c=2$ 
we find that the central charge contribution from the ghost sector is 
$c=4$ ($d=4$), $c=10$ ($d=6$) and $c=22$ ($d=10$) and therefore 
the total central charge vanishes.

A comment is in order here. Since we have shown that the central charge 
vanishes and that the Lorentz current algebra closes, are we to interpret 
the models as critical superstrings in $d<10$? We do not believe this to be 
the correct interpretation. 
Rather, the point of view that is taken in~\cite{Wyllard:2005} is that 
the models should be thought of as the compactification-independent 
sector of the superstring compactified on a Calabi-Yau manifold $CY_{l}$ 
down to $d=10-2l$. 

The zero-mode saturation rules that follow from the above analysis 
takes the schematic form (see \cite{Wyllard:2005} for more details)
\be\label{satrule}
\lb 0|  \la^{n-2} \tha^{ n  } |\Omega \rb \neq 0 \,,
\ee
where (schematically) $|\Omega \rb = \prod_{A}^{q} Y^A |0\rb$ with 
(as in \cite{Berkovits:2004a}) $Y^A = C^A \tha \de(C^A \la)$. 
The integer $q$ in this expression is given 
by $2$ ($d=4$), $5$ ($d=6$) and $11$ ($d=10$). 


\setcounter{equation}{0}
\section{BRST operator, cohomology and vertex operators}
\label{N3}

In this section we initiate a discussion of the BRST structure of the 
lower-dimensional pure-spinor models. For simplicity, we only consider 
the open (or left-moving) sector of the superstring. 

In the $d=10$ pure-spinor formalism, the physical states of 
the superstring are obtained from vertex operators in the cohomology of the 
BRST operator (we only display the left-moving part)
\begin{equation} \label{QBRST}
Q = \oint \lambda^{\alpha} d_{\alpha}\,. 
\end{equation}
The natural guesses for the BRST operators of the lower-dimensional models 
are
\bea\label{QQ}
Q =  \oint \lambda^{\alpha} d_{\alpha}\,, && (d=4) \nonumber \\ 
Q = \oint \lambda^{\alpha}_{I} d^{I}_{\alpha}\,. && (d=6)
\eea 
These operators are nilpotent because of the pure-spinor constraints. 
Notice that a similar operator was used in \cite{Grassi:2004xc}. 
There, the dimensional reduction to $d=4$ preserving $\cN=2,3,4$ supersymmetry 
was studied and the relation with harmonic superspace was shown. 
One of the key points is that the BRST operator obtained after 
the introduction of harmonic variables becomes similar to the above 
operator and only after the introduction of additional constraints 
(analytical constraints) are the equations motions recovered. 
We now analyse the cohomology of the $Q$'s given in (\ref{QQ}) for massless 
states.\footnote{
The solutions to the BRST cohomology equations can also be computed 
using the technique of spinorial cohomology discussed 
in~\cite{Cederwall:2001a,Cederwall:2001b}. In addition, we should 
mention that there is vast literature on solving superspace constraints 
(Bianchi identities) in $d=4$ and $d=6$.} 

In $d=4$ at ghost number zero BRST-closedness requires 
$\la^\al D_\al \Phi =0$. 
If this is to hold for {\it all} ways of solving the pure-spinor constraint 
then $D_\al \Phi = 0$ has to hold; thus $\Phi = \mathrm{const}$ is the only 
solution.
At ghost number one, 
if one works in the Weyl basis, one finds that in order for $Q$ to 
give zero when acting on 
$\la^a A_a(x,\tha) + \bla^{\dot{a}}\bA_{\dot{a}}(x,\tha)$, 
one has to have $A_a = D_a M -D_a N$ and 
$\bA_{\dot{a}} = \bD_{\dot{a}} M + \bD_{\dot{a}} N$. 
Here the terms involving $M$ correspond to a BRST exact piece, 
whereas the field content of $N$ constitute the cohomology and 
is exactly that of (off-shell) 
$\cN=1$, $d=4$ Yang-Mills. Note that $Q$ does {\it not} 
put the theory on-shell; 
it only selects the right (off-shell) field content. 
To put the theory on-shell one needs to add the Virasoro constraint 
(and possibly some other constraints as well; superspace might give a 
clue about the minimal number of constraints one needs). We do not have a 
detailed understanding of how this works. 
In the Dirac basis, at ghost number one, we can write  
\bea\label{verA}
U  = \la^{\al} A_{\al}(x,\tha) \,.
\eea 
The BRST condition $Q U=0$ implies $\la^{\al} \la^{\bet} D_{\la} A_{\bet} =0$. 
Therefore, the most general solution is given by 
\be\label{verB}
U = 
\la^{\al} \left( D_{\al} M + 
(\Ga^{mnpq})_{\al}{}^{\bet} D_{\bet} N_{mnpq} \right) = 
\la^{\al} \left( D_{\al} M + 
(\Ga^{5})_{\al}{}^{\bet} D_{\bet} N_{5} \right) 
\,.
\ee
The above vertex operator (\ref{verA}) is invariant under the gauge 
symmetry $\de U = \{Q , \Om\}$ for a scalar superfield $\Omega$.
Therefore the first term in (\ref{verB}) can be removed 
by a gauge transformation, and the 
second term represents the cohomology. Notice that 
we have to require the reality condition in order not to spoil 
the hermiticity of the theory.\footnote{A remark is in order here. 
In $d=10$ and in lower dimensions, the solution of pure-spinor constraints 
can be only achieved by using complex spinorial 
fields. For example in $d=10$, the pure spinors are only Weyl and 
not Majorana-Weyl. 
However, at the level of the path integral and in all manipulations 
only the field $\la$ appears whereas its complex conjugate is absent. However, 
if the pure-spinor constraints are not solved explicitly, 
one does not see this phenomenon. In $d=4$, the explicit solution of the pure 
spinor constraint breaks the hermiticity properties, but it does not spoil 
the Lorentz invariance of the theory.} 
This implies that the degrees of freedom are represented 
by a real scalar superfield $N_{5}$. At higher ghost number the cohomology 
is empty. This can be understood by computing the zero-momentum cohomology. 
Zero-momentum cohomology is the BRST cohomology 
computed on the space of functionals independent of the spacetime 
coordinates $x^{m}$. On general grounds it turns out that at any ghost number, 
the zero-momentum cohomology is larger than the non-zero one (only 
a subset of the cohomology classes of zero-momentum cohomology can be lifted  
to the non-zero momentum one) and, 
therefore, if the  zero-momentum cohomology is empty at a given 
ghost number, also the corresponding non-zero momentum cohomology 
is empty. In the present case, the zero-momentum cohomology 
is given by the following generators 
\be\label{verC}
1\,,~~~~  \la^{a} \bar\theta^{\dot a} + \bar\la^{\dot a} \tha^{a}\,, ~~~~
 \la^{a} \bar\tha^{2} - \bar\la^{\dot a} \bar\tha_{\dot a} \tha^{a}\,, ~~~~
 \bar\la^{\dot a} \tha^{2} - \la^{a} \tha_{a} \bar\tha^{\dot a}\,, ~~~~
 \bar\la^{\dot a} \bar\tha_{\dot a}\tha^{2} - \la^{a} \tha_{a} \bar\tha^{2}\,. 
\ee
One can also analyse the cohomology at the first massive level. 
This analysis is performed in \cite{Grassi:2005b}.

Integrated vertex operators can be constructed in same way as in $d=10$, 
i.e.~via the descent equation: if ${\cal V}$ is the integrand of 
the integrated vertex operator then (for open superstrings) 
$[Q, {\cal V}] = \partial U$. The massless integrated vertex operator can be written in the Weyl basis as
\be \label{Vint}
V = \oint [ \pa x^m A_m + \pa \tha^a  A_a + \pa \btha^{\dot a} \bA_{\dot a}
+ d_a W^a + \bd_{\dot a} \bW^{\dot a} + \half N^{mn} F_{mn} ] \,, 
\ee
where the decent equation implies that the superfields 
$A_m$, $W^{a}$, $\bW^{\dot a}$ and $F_{mn}$ satisfy the usual 
$d=4$ SYM superspace relations, and $A_a$ and $\bA_{\dot a}$ are as above. 
Plugging these results into the above expression (\ref{Vint}) 
one finds exact agreement 
with the vertex operator in the hybrid superstring \cite{Berkovits:1994a}, 
except for the fact that the term involving $N^{mn}$ is not present 
in the hybrid superstring. This is discussed further in \cite{Wyllard:2005}.

We have not studied the construction of scattering amplitudes using 
the above vertex operators, but we note that from the saturation rule given 
in (\ref{satrule}) and the ghost number of $U$ it seems hard to use the 
same prescription as in $d=10$ \cite{Berkovits:2000a}. 

This concludes our discussion of the massless cohomology in $d=4$. 
But before we move on to the $d=6$ case a comment is in order. 
There is also another way one could compute the cohomology: 
first solve the pure-spinor constraint and plug the solution into $Q$, 
and then compute the BRST cohomology. 
This is rather easy to analyse in the present case. 
As we saw in section \ref{N0}, the pure-spinor constraint is 
solved by choosing either $\la^{a} =0$ or $\la^{\dot a}=0$ 
(at the level of the path integral both solutions have to be 
taken into account; this will be discussed in a forthcoming 
paper~\cite{Grassi:2005b}). By plugging e.g.~the solution 
$\la^{a}=0$ into the BRST formula, one finds the new operator  
$Q'= \oint \la^{\dot a} d_{\dot a}$ which of course is also nilpotent. 
The massless cohomology of this new operator can be straightforwardly 
computed: every chiral superfield $\Phi(y,\theta)$ 
satisfying $D_{\dot a} \Phi=0$ is an element of the cohomology 
at ghost number zero. 
A chiral superfield has $4$ bosonic and $4$ fermionic degrees of freedom. 
It has the same number of independent (off-shell) components as 
the gauge multiplet which occurred in the calculation above and 
indeed they can be mapped into each other. 
(A similar phenomenon is discussed in \cite{Metsaev:2000}.) 
The exact relation between the two cohomologies needs to be understood better.

In $d=6$ the cohomology, at zero ghost number, of the operator in (\ref{QQ}) 
 again contains only a constant.  
At ghost-number one the unintegrated vertex operator can be written 
$U=\la^\al_I A^I_\al$. 
The BRST condition implies $D_{(\al}^{(I} A_{\beta)}^{J)} 
= 0$ which is solved by \cite{Howe:1983} 
\be\label{seiA}
A_{\al}^{J} = D_\al^J M + \ep_{KI} D_\al^I D^{JKLM} N_{LM}\,,
\ee 
where 
$D^{JKLM} = \ep^{\al\beta\ga\de} D_\al^{(J} D_\beta^K D_\ga^L D_\de^{M)}$. 
Here the term involving $M$ is BRST exact, whereas 
it is known~\cite{Howe:1983} that $N_{LM}$ describes $\cN=1$ $d=6$ (off-shell) 
super-Yang-Mills~\cite{Howe:1983}. The $d=6$ cohomology was also computed in 
\cite{Cederwall:2001b} using the (equivalent) spinorial cohomology framework.
As in $d=4$ one can also investigate integrated vertex operators etc. but 
we will not do so here.

Despite the strange nature of the pure spinors in  $d=3$ 
this case can also be analysed using the above method. 
It will now be shown that the cohomology is well defined 
and at the lowest (massless) level describes the gauge supermultiplet. 

The most general vertex operator of ghost number 1 at the massless level is 
\be\label{treA}
U = \la^{\al} A_{\al}(x,\tha)\,,
\ee
which is defined up to the gauge transformation $\delta A_{\al} 
= D_{\al} \Omega$ for some scalar superfield $\Omega$. 
Imposing BRST invariance and using the pure-spinor condition, 
we see that there is no constraint on the superfield 
$A_{\al}$. However, by using the gauge symmetry, we can easily see 
that 
\be\label{treH}
A_{\al} = a_{(\al\bet)}(x) \tha^{\bet} + u_{\al}(x) \tha^{2}\,, ~~~~~
\delta a_{\al\bet}(x) = 
\partial_{\al\bet} \omega(x)\,.
\ee
This is exactly the field content and gauge symmetry 
of super-Yang-Mills in $d=3$. 
It is easy to analyse also the higher ghost-number cohomology 
and it turns out that there no massless cohomology with ghost number 
larger than 1. This poses the problem how to construct antifields, but 
this question is also related to the prescription for the saturation rule 
which is not discussed in this note.  
The computation of the cohomology at the massive levels is left to 
a separate publication \cite{Grassi:2005b}.

An even simpler (albeit somewhat degenerate) case occurs in $d=2$. 
Consider the case of $\cN=(2,0)$ supersymmetry with superspace variables 
$(x^{m}, \theta^\al_I,p_\al^I)$, where $I=1,2$ labels a doublet of 
Majorana-Weyl spinors. We take the pure-spinor constraint satisfied by $\la_I$ 
to be 
\be \label{2d-pure-spinor}
\de^{IJ} \lambda_I \gamma^{m} \lambda_J  = 0\,, \qquad I,J=1,2\,.
\ee 
As in $d=3,4,6$ the only cohomology at ghost number zero 
is a constant. At ghost number one, acting with $Q$ on $\la_I A^I$ one finds 
the constraint $D^{(I}A^{J)} - \half \de^{IJ} D^K A^L \de_{KL} = 0$.
The solution is $A_I = D_I M - \ep_{IJ} D_J N$, or  
$A_1 = D_1 M - D_2 N$ and $A_2 = D_2 M + D_1 N$. Here the piece involving 
$M$ is BRST-trivial, whereas the cohomology is contained in $N$.
More details about this $d=2$ pure-spinor model 
are presented in \cite{Wyllard:2005}.


\section{Open problems and discussion}
\label{N4}

In this note we have summarised some basic facts about 
pure-spinor superstrings in lower dimensions. 
A lot of work can be done and, in our opinion, 
should be done in this direction. Some immediate questions which arise are 
how/if these models are related to compactifications of the RNS superstring, 
and how/if they are related to the so called hybrid 
superstrings~\cite{Berkovits:1994a,Berkovits:1994b} which 
describe compactifications of the superstring with manifest 
super-Poincar\'e covariance in the non-compact directions. Some of these 
questions are touched upon in \cite{Wyllard:2005}.

The new models are simpler than the $d=10$ pure-spinor superstring and 
provide a nice laboratory for studying quantum field theories 
with constraints. One of the most interesting aspects which is not 
yet understood (even for the $d=10$ model) is the origin of the BRST 
operator\footnote{We should mention that it has been argued 
 \cite{Matone:2002} that the pure-spinor formulation can be obtained 
from the superembedding formalism. In addition, in the modified GS action 
given in \cite{Aisaka:2005}, one can see the pure-spinor BRST operator 
emerging. Finally, in \cite{Gaona:2005yw} the BFT embedding is 
used to motivate the BRST operator.}. 

Another interesting question to understand  
is how the Virasoro constraints are implemented in the new framework. 
Let us elaborate on this issue. A basic ingredient of string 
theory is the conformal invariance and the reparametrisation 
invariance on the worldsheet. At tree level a detailed understanding 
is perhaps not really needed, but at higher genus the conformal 
invariance and the Virasoro constraints play a fundamental role. 
Only recently a prescription for higher genus computations has been 
constructed~\cite{Berkovits:2004a}, but it is mainly based on geometrical 
properties and counting of degrees of freedom. 
The role of the Virasoro constraints is not manifest. (The same type 
of problem can be also found in the formulation of higher genus prescriptions 
for topological strings, see for example \cite{Craps:2005wk,Grassi:2004tv}.) 
The formulation of tree level and higher genus computations for 
the new models is therefore very interesting.  

There is another aspect worth mentioning here. As we have seen the 
from the computation of the cohomology, the massless states of these models 
are described by off-shell multiplets of super Yang-Mills. This means that 
some additional constraints should be added in order to describe 
the physical massless and massive degrees of freedom. These additional 
constraints play an important role in the formulation of amplitudes because 
the auxiliary fields and gauge degrees of freedom should drop out from 
correlation functions. Since one of the main differences between the  
new lower-dimensional models and the pure-spinor superstring in $d=10$ 
is that the supersymmetric multiplets are off-shell when $d<10$, 
we hope that from the models presented in this paper the Virasoro constraints 
can be understood. 

\medskip

{\bf Note added:} After this paper was completed, we 
received a copy of \cite{Berkovits:2005c} where pure-spinor 
superstrings in $d=4$ are also discussed. In particular, the role of the 
BRST operator one obtains by explicitly solving the pure-spinor constraint 
in terms of a Weyl spinor (the operator that we called $Q'$ in section 
\ref{N3}) was clarified and a scattering amplitude prescription for the 
corresponding theory was given. 
Furthermore, it was argued that the resulting model is related to 
a chiral sector of superstring theory compactified to  $d=4$ on 
a Calabi-Yau manifold. 


\section*{Acknowledgements}
We would like to thank Francisco Morales, Yaron Oz and Ricardo Schiappa 
for discussions.

\appendix

\setcounter{equation}{0}
\section{Flat superspace in $d=4,6$} \label{appFlat}

\subsection{$\cN=1_{4}$ in $d=4$}
The ${\mathcal{N}}=1$, $d=4$ superspace coordinates are 
in Dirac notation $\left( x^{m}, \theta^{\al}\right)$ ($\al=1,\ldots,4$) 
and in Weyl notation 
$\left( x^{m}, \theta^{a}, \theta^{\dot{a}} \right)$ 
($a=1,2$, $\dot{a}=1,2$). 
We denote the $4{\times}4$ gamma matrices acting on Dirac spinors 
by $\Gamma^m$. The charge conjugation matrix used to raise and lower 
indexes will not be written explicitly.
 
The supersymmetry transformations acting on superfields 
are generated by (in Dirac notation)
\begin{equation}
\mathcal{Q}_{\alpha} = \frac{\partial}{\partial \theta^{\alpha}} -
\frac{1}{2}  \left( \Ga^{m} \tha \right)_{\alpha}
\frac{\partial}{\partial x^{m}}\,, 
\end{equation}
satisfying   
\begin{equation}
\left\{ \mathcal{Q}_{\alpha}, \mathcal{Q}_{\beta} \right\} = -
\Ga^{m}_{\alpha\beta} \frac{\partial}{\partial x^{m}}\,.
\end{equation}
The vector field $\frac{\partial}{\partial x^{m}}$ is invariant under the
supersymmetry transformations, as is the usual supersymmetric derivative,
\begin{equation}
D_{\alpha} = \frac{\partial}{\partial \theta^{\alpha}} +
\frac{1}{2}  \left( \Ga^{m} \tha \right)_{\alpha}
\frac{\partial}{\partial x^{m}}\,, 
\end{equation}
which satisfies,
\begin{equation}
\left\{ D_{\alpha}, D_{\beta} \right\} = \Ga^{m}_{\alpha\beta}
\frac{\partial}{\partial x^{m}}\,.
\end{equation}

\subsection{$\cN=(1,0)_{8}$ in $d=6$}

The ${\mathcal{N}}=(1,0)$, $d=6$ superspace coordinates are 
$\left( x^{m}, \theta_I^{\alpha} \right)$ ($\al=1,\ldots,4$, $I=1,2$). 
The supersymmetry transformations acting on superfields 
are generated by
\begin{equation}
\mathcal{Q}^I_{\alpha} = \frac{\partial}{\partial \theta_I^{\alpha}} -
\frac{1}{2} \ep^{IJ}  \left( \gamma^{m} \theta_J \right)_{\alpha}
\frac{\partial}{\partial x^{m}}\,,
\end{equation}
satisfying
\begin{equation}
\left\{ \mathcal{Q}^I_{\alpha}, \mathcal{Q}^J_{\beta} \right\} = -\ep^{IJ}
\gamma^{m}_{\alpha\beta} \frac{\partial}{\partial x^{m}}\,.
\end{equation}
The vector field $\frac{\partial}{\partial x^{m}}$ is invariant under the
supersymmetry transformations, as is the usual supersymmetric derivative,
\begin{equation}
D^I_{\alpha} = \frac{\partial}{\partial \theta_I^{\alpha}} 
+ \frac{1}{2} \ep^{IJ} \left( 
\gamma^{m} \theta_J \right)_{\alpha} \frac{\partial}{\partial x^{m}}\,,
\end{equation}
which satisfies,
\begin{equation}
\left\{ D^I_{\alpha}, D^J_{\beta} \right\} = \ep^{IJ} \gamma^{m}_{\alpha\beta}
\frac{\partial}{\partial x^{m}}\,.
\end{equation}

\begingroup\raggedright\endgroup

\end{document}